# High-altitude Magnetospheric Emissions from Two Pulsars

Mao Yuan[1,2], Weiwei Zhu[1], Michael Kramer[3], Bo Peng[1], Jiguang Lu[1], Renxin Xu[4,5], Lijing Shao[5], Hong-Guang Wang[6], Lingqi Meng[1,2], Jiarui Niu[1,2], Rushuang Zhao[1], Chenchen Miao[1,2], Xueli Miao[1], Mengyao Xue[1], Yi Feng[7], Pei Wang[1], Di Li[1], Chengming Zhang[1], David J. Champion[3], Emmanuel Fonseca[8,9], Huanchen Hu[3], Jumei Yao[10], Paulo C. C. Freire[3], and Yanjun Guo[3]

[1] National Astronomical Observatories, Chinese Academy of Sciences, Beijing 100101, People's Republic of China; yuanmao@bao.ac.cn, zhuww@nao.cas.cn, pb@bao.ac.cn
[2] College of Astronomy and Space Sciences, University of Chinese Academy of Sciences, Beijing 100049, People's Republic of China
[3] Max-Planck-Institut für Radioastronomie, Auf dem Hügel 69, D-53121 Bonn, Germany; michael@mpifr-bonn.mpg.de
[4] Department of Astronomy, Peking University, Beijing 100871, People's Republic of China
[5] Kavli Institute for Astronomy and Astrophysics, Peking University, Beijing 100871, People's Republic of China
[6] Department of Astronomy, Guangzhou University, Guangzhou 510006, People's Republic of China
[7] Research Institute of Artificial Intelligence, Zhejiang Lab, Hangzhou, Zhejiang 311121, People's Republic of China
[8] Department of Physics and Astronomy, West Virginia University, P.O. Box 6315, Morgantown, WV 26506, USA
[9] Center for Gravitational Waves and Cosmology, West Virginia University, Chestnut Ridge Research Building, Morgantown, WV 26505, USA
[10] Xinjiang Astronomical Observatory, Chinese Academy of Sciences, 150, Science 1-Street, Urumqi, Xinjiang 830011, People's Republic of China

## Abstract

We discover three new weak pulse components in two known pulsars, one in PSR J0304+1932 and two in PSR J1518+4904. These components are emitted about halfway between the main emission beam and the interpulse beam (beam from the opposite pole). They are separated from their main pulse peak by $99^{\circ} \pm 3^{\circ}$ for J0304+1932 and $123^{\circ}.6 \pm 0^{\circ}.7$ (leading) and $93^{\circ} \pm 0^{\circ}.4$ (trailing) for J1518+4904. Their peak-intensity ratios to main pulses are $\sim 0.06\%$ for J0304+1932 and $\sim$0.17% and $\sim$0.83% for J1518+4904. We also analyzed the flux fluctuations and profile variations of the emissions for the two pulsars. The results show correlations between the weak pulses and their main pulses, indicating that these emissions come from the same pole. We estimated the emission altitude of these weak pulses and derived a height of about half of the pulsar's light-cylinder radius. These pulse components are a unique sample of high-altitude emissions from pulsars, and challenge the current pulsar emission models.

## 1. Introduction

The consensual picture of a radio pulsar is a spinning neutron star with a strong dipole magnetic field. In this picture, radio radiation is emitted from a narrow polar beam and manifests as periodic pulses. For most pulsars, the pulse only spans a small range of spin phases, leaving the other region as a "pulse-off" phase.

Other than the main pulse (MP), some pulsars also have an additional interpulse (IP). IPs separate usually 180° from the MPs and are thought to be caused by the opposing emission beam of the MP (Hankins & Fowler 1986; Taylor et al. 1993; Maciesiak et al. 2011). The intensities of IPs reported are often much weaker than those of the MPs, with ratios above 0.5% (Kramer et al. 1998; Kloumann & Rankin 2010; Dai et al. 2015; Sobey et al. 2015; Arzamasskiy et al. 2017; Johnston & Kramer 2019).

Different from IPs, some weak emission components were found much closer to the MP in recycled pulsars (i.e., millisecond pulsars, MSPs; Kramer et al. 1998, 1999; Dai et al. 2015; Gentile et al. 2018; Wahl et al. 2022). Such components are called "weak" components (Dai et al. 2015) and "microcomponents" (Gentile et al. 2018; Wahl et al. 2022). They are characterized by having a much lower strength compared to the MP. Specifically, "microcomponents" are defined as emission components whose strength is <3% of the highest peak (Wahl et al. 2022). Most "weak" components or "microcomponents" in recycled pulsars appear to be extensions of broad MPs (see Wahl et al. 2022), and some are weak IPs, e.g., the case in PSR J1909-3744 (Dai et al. 2015; Wahl et al. 2022). The components (except IPs) in Kramer et al. (1998), Dai et al. (2015), Gentile et al. (2018), and Wahl et al. (2022) are linked to the MPs as a part of complex profiles spanning >100° in phase. Such a complex emission profile is understandable for recycled pulsars as their light cylinders are smaller and their magnetospheres no longer have the shape of a simple dipole due to rapid rotation (Chen & Ruderman 1993; Romani 1996; Kramer et al. 1998; Xilouris et al. 1998; Wahl et al. 2022).

A few special weak-component cases exist that are not fast-rotating pulsars but also exhibit complex pulse profiles that span a large duty cycle, e.g., PSR B0826-34 (Biggs et al. 1985; Gupta et al. 2004) and J1107-5907 (Young et al. 2014). Such pulsars are believed to be aligned rotators (Malov & Nikitina 2013), their emissions are visible because of their special viewing geometry.

In this paper, we discovered three separate weak pulse components with significant polarization by using the Five-hundred-meter Aperture Spherical radio Telescope (FAST). These weak emissions are from two known pulsars. One is from J0304+1932, a bright pulsar with a spin period of 1.4 s. The separation of this weak pulse component is $99^{\circ}.2$, and the amplitude ratio to the MP is $\sim$0.06%. The other two are from J1518+4904, a 40.9 MSP in a double neutron star system. The

**Table 1**
Observation Information For the Two Pulsars

| Pulsar | Obs MJD | Obs Duration | Single Pulse Number | Frequency Band | Channels | Time Resolution |
|---|---|---|---|---|---|---|
| J0304+1932 | 59242 | 0.6 hr | 1171 | 1.0–1.5 GHz | 4096 | 49.152 $\mu$s |
| J1518+4904 | 59604 | 2 hr | 158,465 | 1.0–1.5 GHz | 4096 | 49.152 $\mu$s |

**Note.** We conducted noise injection at the beginning of each observation for polarimetric calibration. The duration of the modulated signal is 1 min.

first component leads the MP by 124°.3 and the intensity ratio IP/MP is ∼0.17%. The other trails the MP by 92°.6 with IP/MP ∼0.83%. In addition, we also observed a bridge emission that connects the MP and these two components of J1518+4904.

These pulse components differ fundamentally from the IPs because of their about 90° separation from their MPs. In terms of strength, they correspond to the concept of "microcomponents" (Gentile et al. 2018; Wahl et al. 2022). However, they do not originate from rapidly rotating pulsars or aligned rotators. Also, these components and their MPs do not have large duty cycles in phase. It appears that these pulses are not extensions of their MPs. These make them unique examples of possible high-altitude emissions from pulsars. We named such separated pulse components orthogonal pulses (OPs).

We performed polarimetric observations on these two pulsars and obtained clear polarization profiles for both the MPs and OPs. Polarimetric analyses are an effective method to recover the emission geometry of a radio pulsar. On the one hand, the linear polarization position angle (PA) helps to determine the emission geometry of a pulsar (Blaskiewicz et al. 1991; Everett & Weisberg 2001; Johnston & Weisberg 2006; Kramer & Johnston 2008; Keith et al. 2010), especially at the origin of separated pulse components such as IPs (Kramer & Johnston 2008; Keith et al. 2010). On the other hand, emission components with different polarization degrees usually have different emission heights or regions in the magnetosphere where pulses propagate in different ways toward the line of sight (LOS; Lyne & Manchester 1988; Radhakrishnan & Rankin 1990; Rankin 1990; Manchester 1995; Weltevrede & Johnston 2008; Johnston & Kramer 2019). For example, IPs from the edge of the radiation beam are usually highly linearly polarized (Johnston & Kramer 2019).

In this article, we present our observations in Section 2 and our analysis in Section 3. We present a discussion in Section 4. At last we summarize our conclusions in Section 5.

## 2. Observations and Data Processing

We used the central beam of the 19-beam receiver of FAST to observe the two pulsars and recorded the data as a time–frequency data set in 8-bit PSRFITS format (Hotan et al. 2004) with four polarizations. The time and frequency sampling resolutions were 49.152 $\mu$s and 0.122 MHz, respectively. The full bandwidth is 1.0–1.5 GHz.

We conducted multiple observations of these two pulsars to make sure the OPs are a real pulsar signal. We observed J0304+1932 twice. Since we have other science proposals for J1518+4904, we observed it several times. For each observation, we process the data using the PSRCHIVE software (Hotan et al. 2004), including zapping RFI channels (by hand using the `pazi` function), baseline removal, and polarimetric calibration. The noise signal for calibration is modulated with a period of 0.201326592 s and a duration of 1 min. Since we have not performed observations on calibrated sources, there is no absolute flux calibration in this work.

All of our observations on these two pulsars yield significant pulse signals at the OP phase across the entire frequency band. We chose the observation with the highest signal-to-noise ratio (S/N) for analysis. The information about the two observations is presented in Table 1. The integration time of these two observations was 0.6 hr (J0304+1932) and 2 hr (J1518+4904), including a setup time of 10 minutes.

We used the PSRCHIVE package `pdv` to derive single pulses from the observations in Table 1. We excised baseline-unstable single pulses for both pulsars. The strategy is to select samples whose standard deviation (STD) of the MP-off phase exceeds the $\pm 3\sigma$ range of the STD statistics for all single pulses. We define the MP-off phase as the region in the mean profile with an intensity less than 1% of the highest peak. We completely removed 117 samples from the single pulse of J0304+1932 and 31,626 samples from the single pulse of J1518+4904. Since J0304+1932 is a nulling pulsar (Herfindal & Rankin 2009), the samples used for RFI mitigation contained only single pulses with detectable MP emission. A further introduction to this nulling can be found in Section 3.1.2.

## 3. Results

In this section, we show the profile observed with FAST first. Then we present our analysis of the OPs.

We present the calibrated integrated pulse profile and the two-dimensional frequency–phase map of the two pulsars in Figure 1. In particular, PSR J0304+1932 is a nulling pulsar (Herfindal & Rankin 2009), and we present the pulse-on and pulse-off phases of the single pulses we observed (see in Table 1) in Figure 2.

Because the two pulsars differ in intensity and emission properties, we separately analyzed their emission geometries. OP emission heights are estimated at the end of this section.

### *3.1. J0304+1932*

In this section, we study the profile and polarization in Section 3.1.1, and then we analyze the nulling phenomenon in Section 3.1.2.

#### *3.1.1. Integrated Profiles and Polarization PA*

PSR J0304+1932's MP has two peaks, consistent with a typical cone emission profile (Gil et al. 1993; Rankin 1993; Kramer et al. 1994; Young & Rankin 2012). We zoom in on the mean profile as well as the polarization profile in Figure 3 and see that the OP is also double peaked with a high degree of linear polarization. The peak intensity of OP is only ∼0.06% of the MP. We measure the pulse widths of the MP and OP as shown in Table 2.

The polarization PA swing is shown in the top panels of Figure 3. The PA of the OP is flat, while the PA of the MP has a swing with a typical "S" shape. The PA of the MP has been observed by other telescopes and fitted with a rotating vector model (RVM; Radhakrishnan & Cooke 1969; Rankin 1993; Everett & Weisberg 2001), resulting in the following geometric parameters in the L band. The angle $\alpha$ between the magnetic and rotation axis is 110°–160°; the angle $\beta$ between the rotation axis and LOS is $\sim$0° (Blaskiewicz et al. 1991). It is clearly not an aligned rotator.

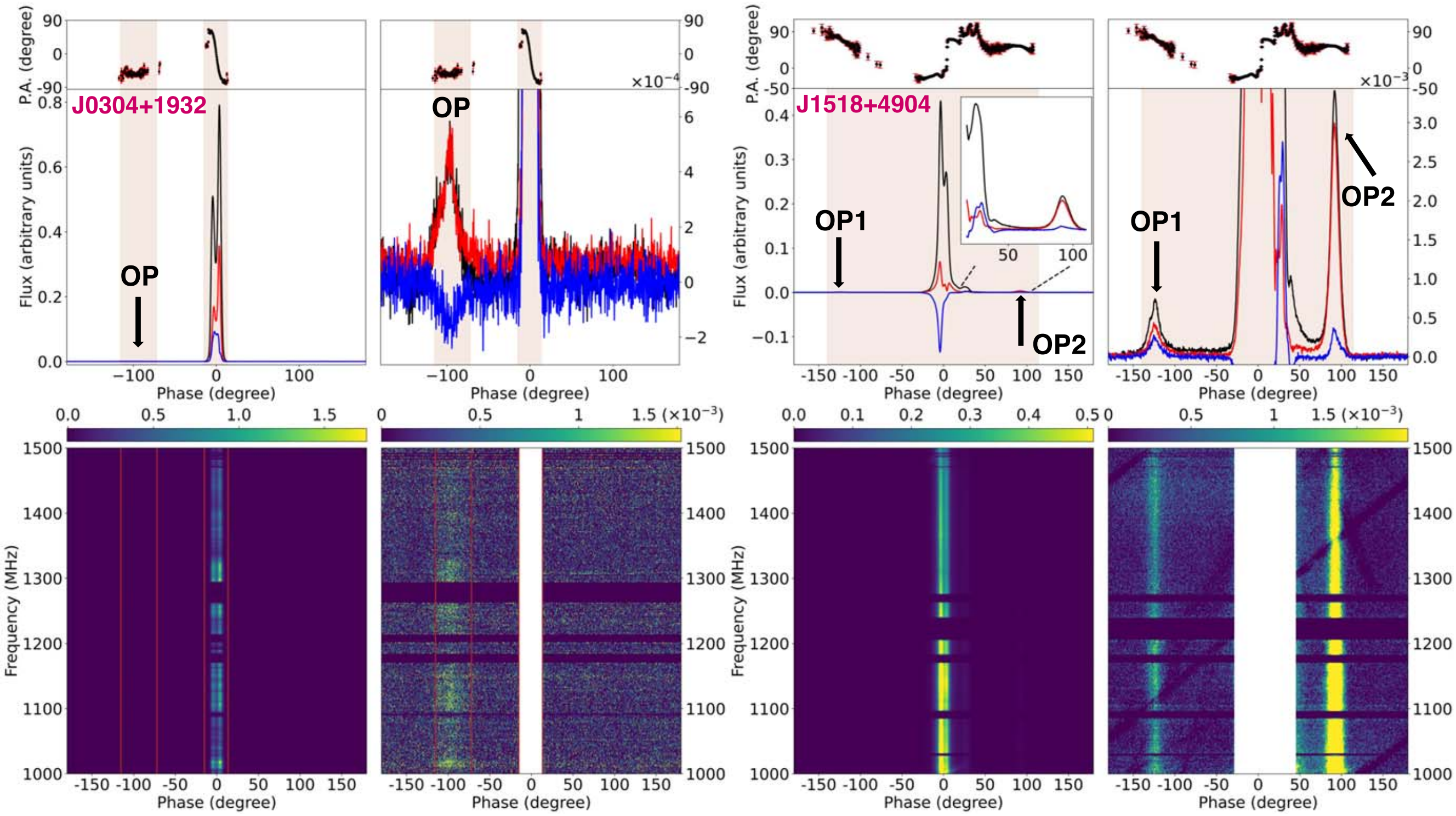

**Figure 1.** The pulse profile (including the PAs) of the two pulsars. On the top panels, the black, red, and blue lines correspond, respectively, to the total intensity, the linear, and the circular polarization. The shadow areas in these panels refer to the emission zone that is over $1\sigma$ of the pulse-off phase. For J0304+1932, the OP is roughly at $-99^\circ$ in the phase. Its peak intensity is $\sim$0.06% of the MP. For J1518+4904, OP1 (left) is around $-124^\circ$ and OP2 (right) is around 92°. The peak-intensity ratios of the OPs to MP are $\sim$0.17% for OP1 and $\sim$0.83% for OP2. To the right of the MP, there is a postcursor (PC) component at 23°, which has been identified by Kramer et al. (1998). The dark oblique line in the frequency–time map of J1518+4904 is caused by weak interference.

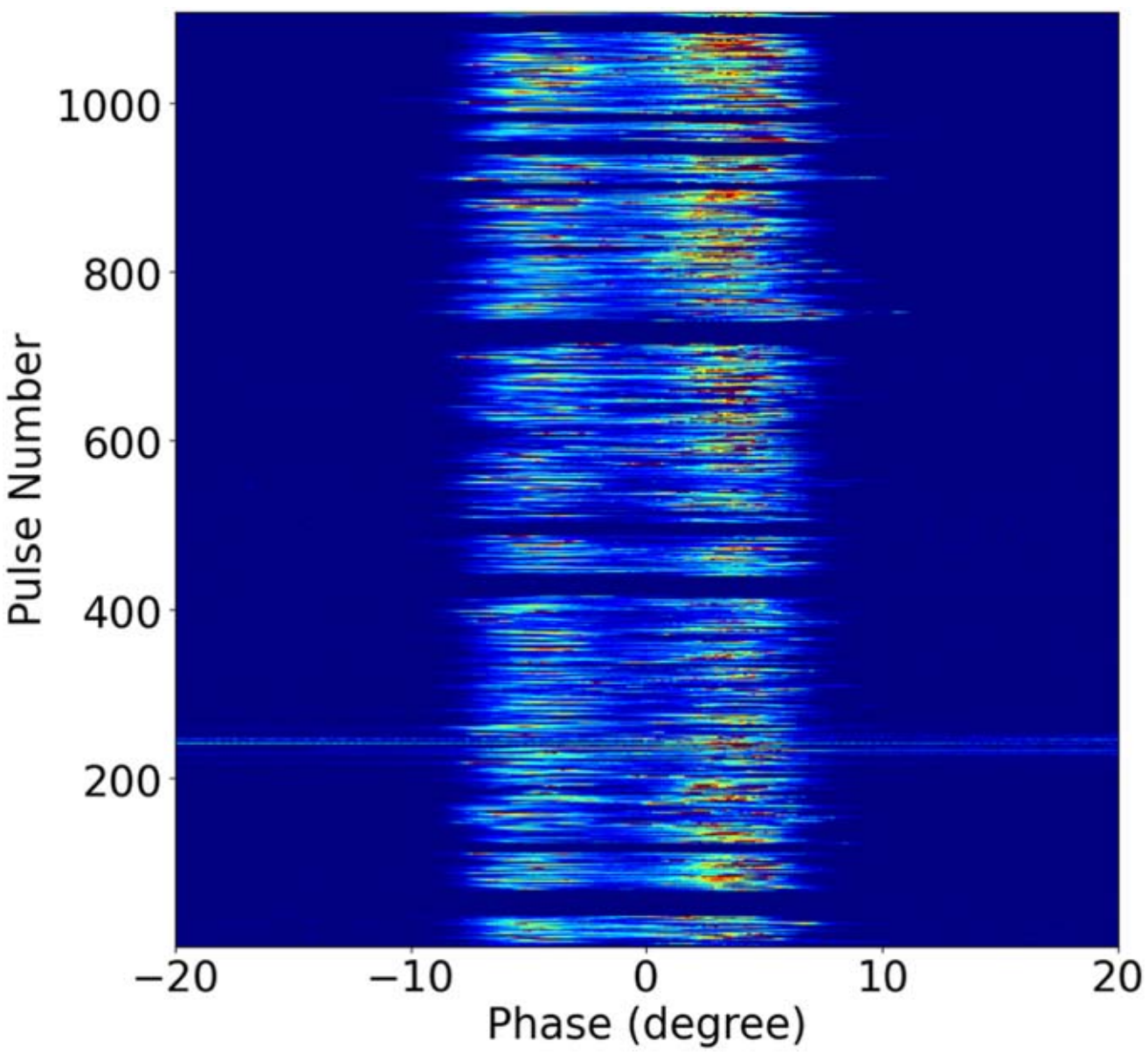

**Figure 2.** The single-pulse sequence of the nulling pulsar J0304+1932. Pulses between 220 and 260 are greatly contaminated by RFI.

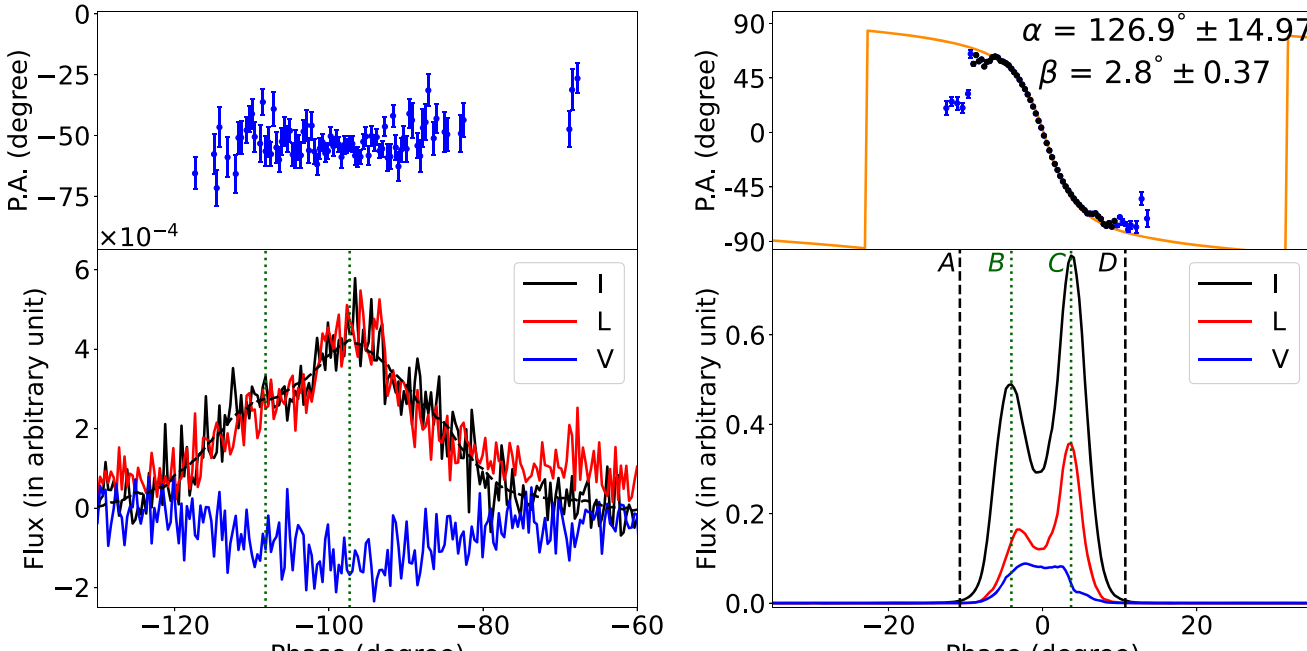

**Figure 3.** Mean profile and PA of J0304+1932. The black, red, and blue lines ($I$, $L$, and $V$) correspond, respectively, to the total intensity, the linear, and the circular polarization. The orange line in the upper panel is the RVM fitted to the PAs. We only fitted PAs with errors $<1^\circ$, which are masked in black. The PAs in blue are ignored during the fit. The green dashed lines in the left panel locate the peaks of the smoothed OP profile. The dashed lines, "$A$" and "$D$" in the lower right panel, are the edges of the MP region defined according to 1% of its peak intensity. The dotted lines, "$B$" and "$C$," respectively locate the longitude of the left and right peaks. The horizontal span of each panel is 80°.

We fit the FAST-detected PA of the MP with the RVM as well. The PA under the model can be described by

$$\mathrm{PA} = \mathrm{PA}_0 + \arctan\left(\frac{\sin\alpha\sin(\phi - \phi_0)}{\sin\zeta\cos\alpha - \cos\zeta\sin\alpha\cos(\phi - \phi_0)}\right)$$

where $\zeta = \alpha + \beta$, $\phi$ is the phase longitude, and $\mathrm{PA}_0$ and $\phi_0$ are, respectively, the PA and phase longitude in the fiducial plane.

**Table 2**
Pulse Width and Polarization Degree of J0304+1932

| | Pulse Width | | $L/I$ | | $V/I$ | |
|---|---|---|---|---|---|---|
| | $W_{10}$(°) | $W_{50}$(°) | $(W_{10})$ | $(W_{50})$ | $(W_{10})$ | $(W_{50})$ |
| OP | ⋯ | 23.7 ± 10.7 | ⋯ | 93.0 ± 6.4% | ⋯ | 33.8 ± 5.5% |
| MP | 15.1 | 11.3 | 14.7% | 15.7% | 1.8% | 1.9% |

**Note.** Because the S/N of OP at 10% peak intensity is below the $3\sigma$ level, we ignore the $W_{10}$ of the OP here. The case of J1518+4904, shown in Table 3 is the same. We ignore measurement uncertainties below 0.1%.

We got the main geometrical parameters as follows: $\alpha = 127° \pm 15°$ and $\beta = 2\overset{\circ}{.}8 \pm 0\overset{\circ}{.}4$. The resulting RVM is shown in the top panels of Figure 3.

We present the three-dimensional radiation geometry of PSR J0304+1932 in Figure 4 based on our fitted parameters. For ease of drawing, we set $\alpha = 127°$ and $\beta = 0°$. In this case, the LOS coincides with the magnetic axis. In this illustration, the rotation axis is set upward vertically, and the trajectory of the LOS forms a cyan circle ⊙O to the bottom of the figure. The neutron star is marked as a red star and the letter "*P*." Symmetric dipole magnetic field lines coming out of the neutron star are presented as blue arcs. The pink cone represents the emission cone of the MP. In this figure, we demonstrate that the emission from the high-altitude magnetosphere could cross the LOS in the correct phase and produce the pulse emission of the observed width. The figure is drawn to scale using the aforementioned geometrical parameters, which includes the opening angle of the MP beam and the orthogonal emission components.

In Figure 5, we show the projection of the magnetic field lines along with the LOS. This allows us to visualize how the observed polarization PA varies with the moving LOS. One can see that when the LOS crosses the center of the beam (MP) the magnetic field lines would be swinging over 180°, and so are the PAs. When the LOS moves across the OP, the magnetic field lines are nearly parallel, and so are the PAs.

In summary, the emission and polarization geometry both strongly support the MP and the OP originating from the same pole and the OP coming from high-magnetosphere-altitude.

### 3.1.2. Temporal Correlation and Nulling

PSR J0304+1932 exhibits a pulse-nulling phenomenon. In this section, we analyze the flux correlation between the MP and OP while the pulsar went in and out of the nulling phase (Figure 2 presents the pulse-nulling phenomenon). We determined the pulse-on and pulse-off single pulses according to the S/N of the MP. The S/N is defined as the value of the average of the on-pulse flux divided by the STD of the off-pulse profile. We choose single pulses with $S/N < 3\sigma$ as nulling pulses. After excising some RFI-contaminated pulses (see Section 2), we collected 179 pulses without detectable signals (nulling pulses). We identify the remaining 992 series normally emitting pulses.

In the top panel of Figure 6, we present the summed profiles of all nulling-phase pulses for both the MP (left) and the OP (right). One can see that there are negligible signals in either of them. In view of the background noise, we cautiously interpret this pulse set as nulling phases. In the following six panels of Figure 6, we present the summed profile of the rest of the normally emitting MP and OP. Each of the first five panels contains 179 pulses, the same as the top panel, and the last panel contains only 96 pulses. In all of these panels of normal mode, clear signals were detected from both the MP and the OP.

This is the second pulsar case in which synchronous nulling occurs between separate pulse components. A similar nulling phenomenon occurs in PSR J1946+1805 (B1944+17), whose MP nulls synchronously with its IP (Hankins & Fowler 1986). Interestingly, this pulsar is believed to be an aligned rotator (Kloumann & Rankin 2010). The synchronous nulling of MP and IP is easily understandable in such an aligned geometry. A counterexample is PSR J1825-0935 (B1822-09). Its IP is null in anticorrelation with the precursor component of the MP while the MP always emits (Hankins & Fowler 1986; Gil et al. 1994). Research shows that it is an orthogonal rotator (Backus et al. 2010; Hermsen et al. 2017).

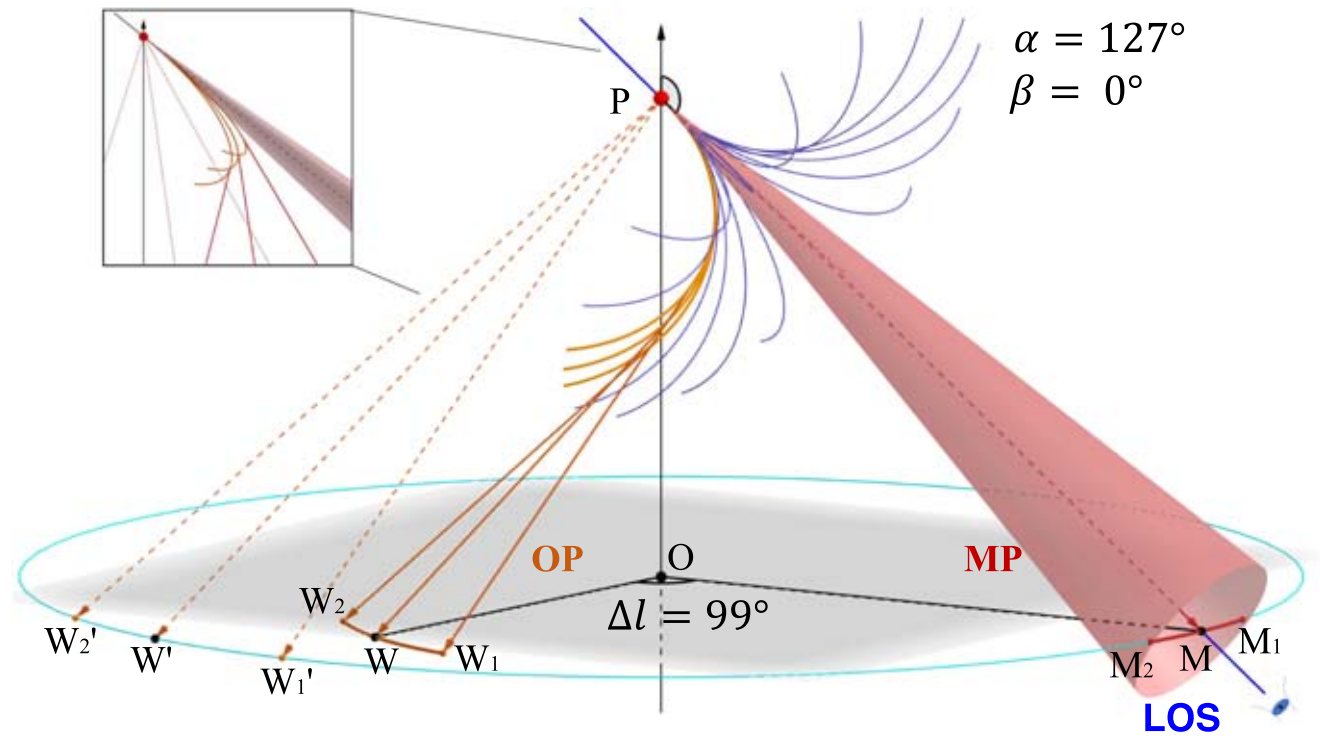

**Figure 4.** Three-dimensional geometric schematic of J0304+1932. $\vec{M}$ and $\vec{W}$ are vectors indicating the emission directions of the MP and OP, respectively. The arcs $\widehat{M_1M_2}$ and $\widehat{W_1W_2}$ are the projected observable emission ranges of the MP and OP, respectively. The widths of these two arcs are, respectively, $\widehat{W_1W_2} = 27°$ and $\widehat{M_1M_2} = 11°$. $\Delta l$ represents the phase separation between the OP and MP.

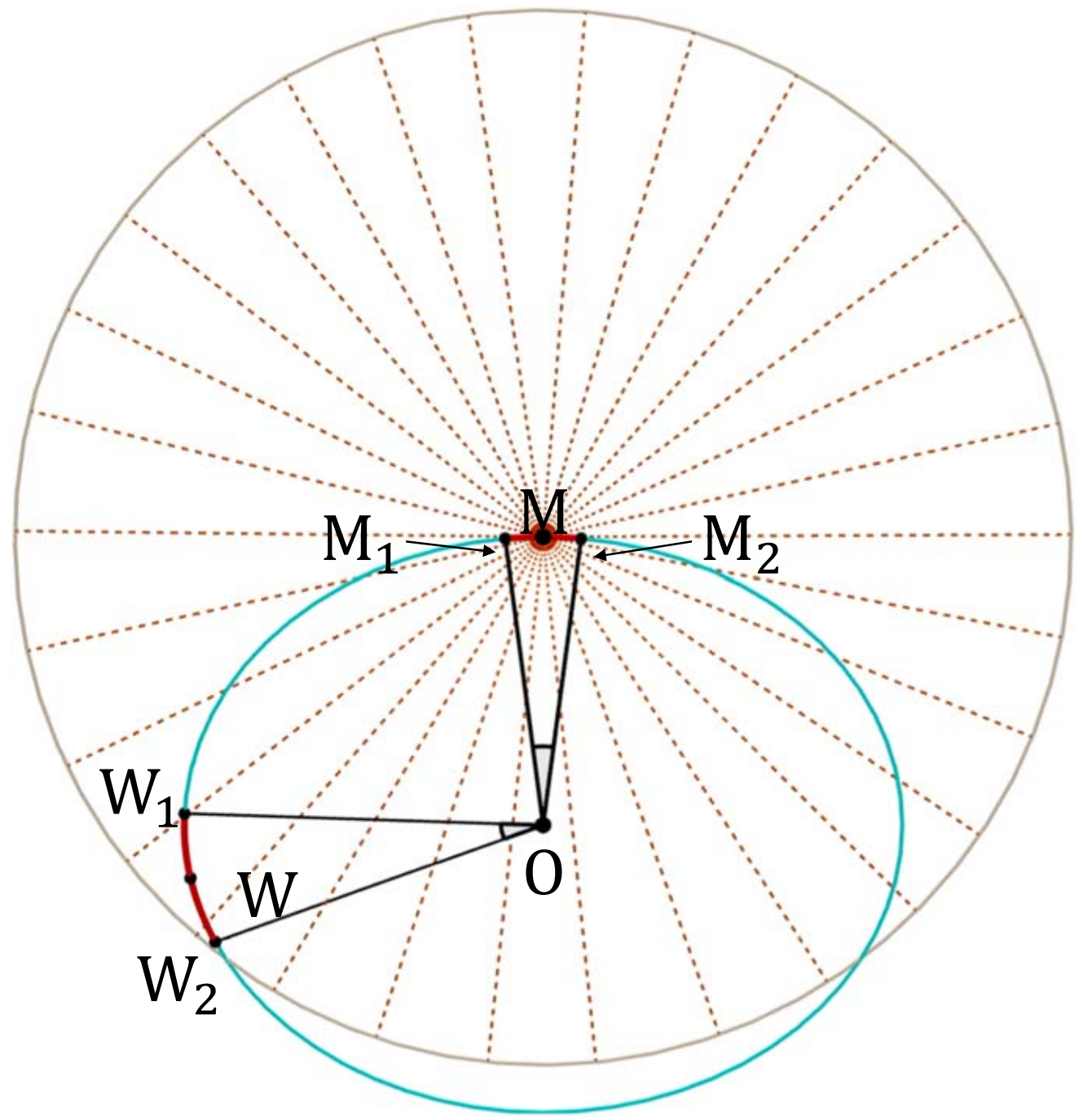

**Figure 5.** A polar projected view of the magnetic axis of PSR J0304+1932, along with the projected trajectory of the LOS. This figure is a two-dimensional projection from Figure 4.

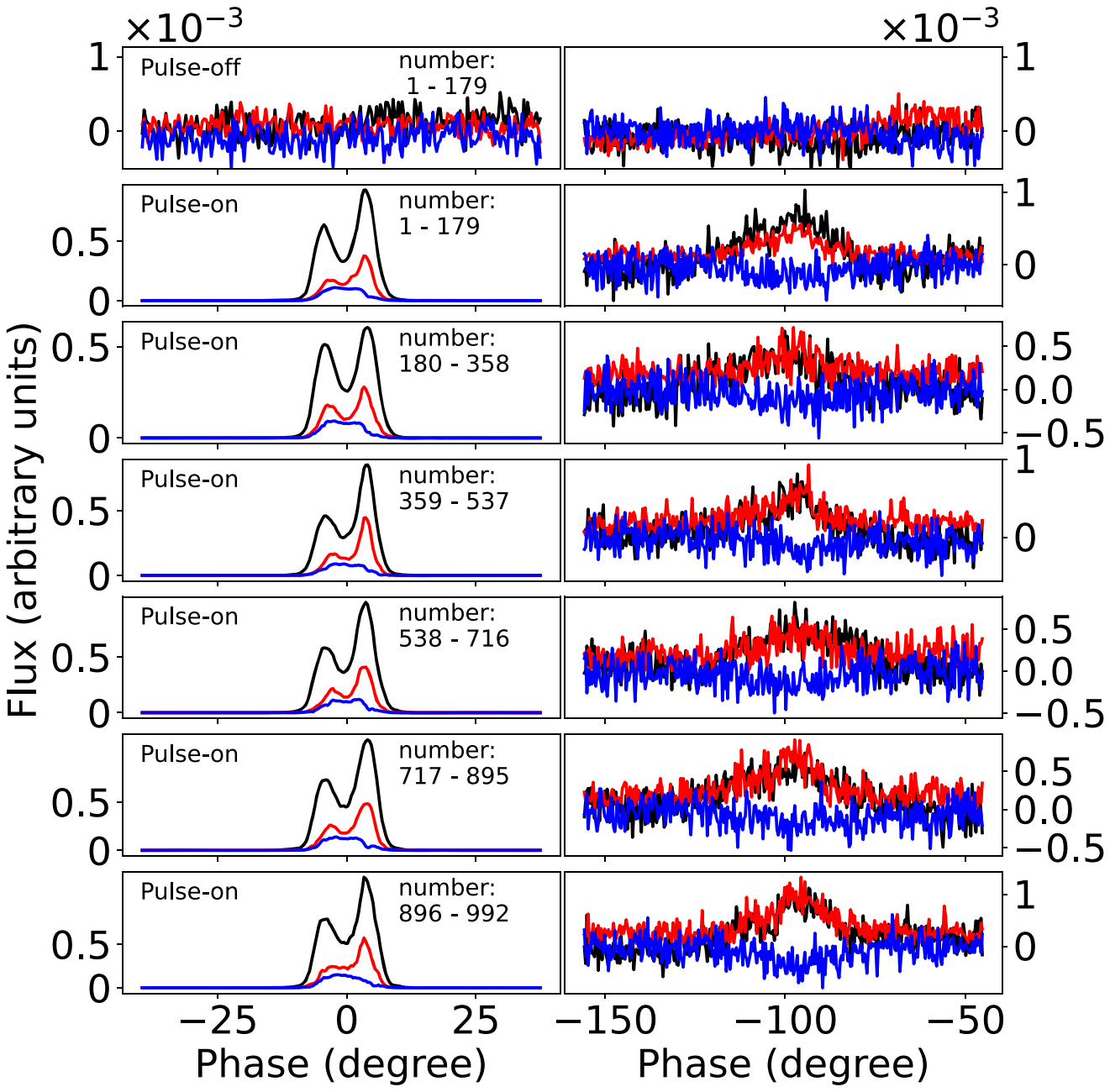

**Figure 6.** The pulse-on and pulse-off phase of J0304+1932. The black, red, and blue lines (*I*, *L*, and *V*) correspond, respectively, to the total intensity, the linear, and the circular polarization. The top subplots are the subintegrated profile of a nulling-pulse subset. The others are profiles of pulse-on subsets.

Pulsar nulling is likely the result of changes in the current distribution in the magnetosphere, i.e., the current is turned off or redistributed (Rankin 1986; Wang et al. 2007; Timokhin 2010). The cases of J1946+1805 and J1825-0935 show that the current in one pole can change globally or partially, but need not switch off synchronously.

Accordingly, the synchronous nulling between the OP and MP of J0304+1932 strongly suggests that both components originate from the same stream of plasma at the same pole. This is consistent with the polarization analysis above.

### 3.2. J1518+4904

In this section, we constrain the origin of the OPs in PSR J1518+4904. The PA swing of J1518+4904 is challenging to fit by the RVM, so we first analyze the mean profile and the PA in Section 3.2.1, then we attempt to reveal the OPs' origin by analyzing their correlations with their MP in Section 3.2.2. Finally, we investigate the profile variations of the MP and OPs in Section 3.2.3.

#### 3.2.1. Integrated Profiles and Polarization Angle

The new emission components of J1518+4904 contain two OPs (OP1 and OP2) and bridge emissions between the OPs and their centered MP. OP1 shows two peaks, while OP2 is a single peak pulse but shows substructures (see Section 3.2.3 for details). The bridge emissions come between the centered MP and the lateral OPs. A PC to the MP was previously identified by Kramer et al. (1998) at 1.4 GHz by using the Effelsberg 100 m radio telescope. Detailed intensity profiles and the PA observed with FAST are shown in Figure 7.

Both OP1 and OP2 are significantly linearly polarized, especially OP2, in which $L/I$ is over 80%. We measure the polarization degree of these emission components together with the pulse width, as shown in Table 3.

J1518+4904 is a mildly recycled pulsar (10–100 ms; O'Shaughnessy & Kim 2010) with a complex polarization profile. Such polarization is usually beyond the transfer predicted by the RVM. An alternative approach is to ignore PAs with jumps or complex variations in phase longitude (e.g., see Everett & Weisberg 2001; Johnston & Kramer 2019). We attempt to fit the PA of J1518+4904 by ignoring parts of the PA (Figure 8). The results are $\alpha = 40.^\circ9 \pm 2.^\circ0$ and $\beta = 29.^\circ3 \pm 4.^\circ4$. Nevertheless, such a fitting result is biased due to the complex PA swing and orthogonal mode jumps, and we do not consider this solution to be reliable.

#### 3.2.2. Correlation in Phase-locked Intensity

We studied the longitude–longitude cross-correlation of the single pulses, to reveal any phase-dependent relations of the OPs and MP. By calculating the correlation coefficient of any two bins in the phase, we can get a two-dimensional longitude–longitude correlation map. For example, the correlation between bins $i$ and $j$ is measured by the covariance $cov(I_i, I_j)$, which $I_{i(\mathrm{or}\ j)}$ is the time-dependent strength series in phase bin $i$ (or $j$). Such a method has widely been used to examine the flux correlation between different emission components, e.g., the MP and IP (Weltevrede et al. 2007; Latham et al. 2012; Kou et al. 2021). Since OP1 is too faint to get credible correlation coefficients, we only present the calculation results from $-30^\circ$ to $110^\circ$ in the phase, shown in Figure 9.

This map clearly shows that OP2 is indeed associated with the MP in strength fluctuation. In detail, it shows an apparent positive correlation between OP2 and the MP.

#### 3.2.3. Correlation in Profile Shape

In this section, we investigate the correlation between the pulse shapes of the MP and OPs. Because OP2 of J1518+4904 was detectable only after integrating hundreds of single pulses, we examine the average profile of the single-pulse series. Nevertheless, the average MP profile is stable after integrating hundreds of pulses arranged by the arrival time. But we can detect shape changes between single pulses, and the changes mainly reflect differences in the strengths between subcomponents of MP. Thus we can detect the averaged shape changes of pulses arranged by MP subcomponent intensity, rather than by the arrival time.

We divided the double-peak MP profile into three subcomponents: the leading component (left), the central component (middle), and the trailing component (right). In this way, all single pulses can be divided into three shape groups, each dominated by one of the three subcomponents. We can then obtain the mean profile for each shape group and analyze the correlation of the shape changes between the MP and OPs.

To define the shape groups, we used two different shape parameters: (1) we calculate the ratio between the integrated

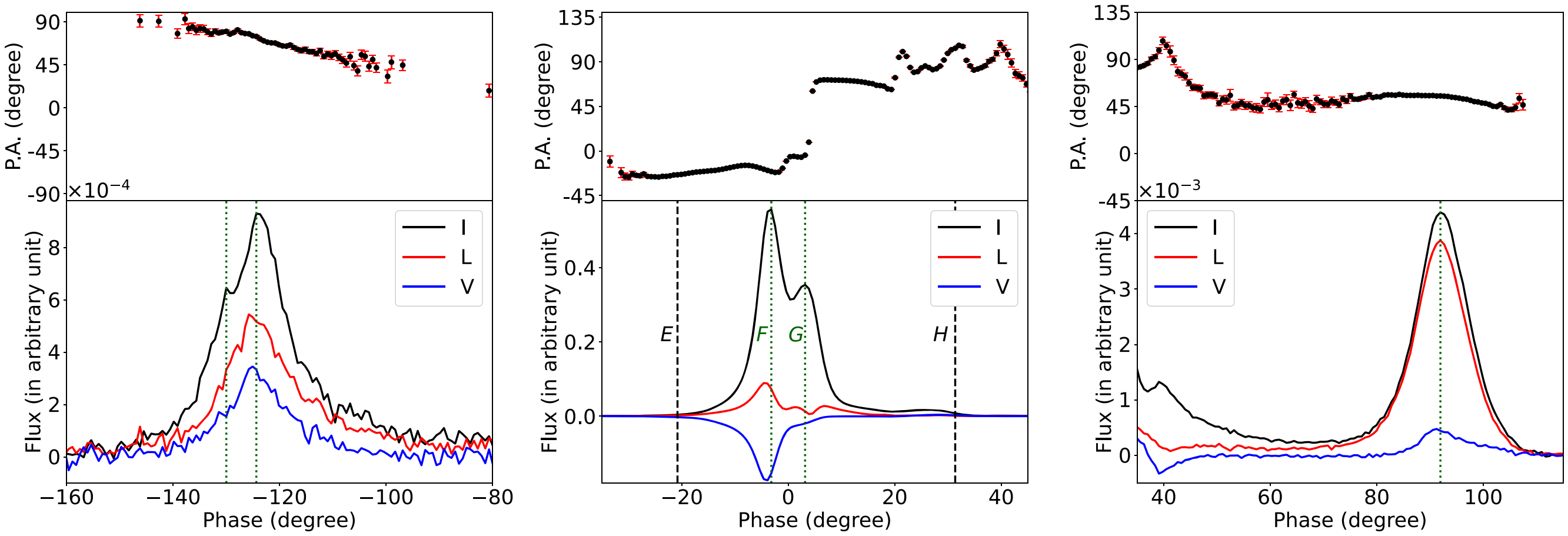

**Figure 7.** Mean profile and the PA of J1518+4904. The black, red, and blue lines ($I$, $L$, and $V$) correspond, respectively, to the total intensity, the linear, and the circular polarization. The left and right panels also show bridge emissions connecting the OPs and MPs. The green dashed lines in the two panels indicate the peaks of the OP profiles. In the central panel, the black dashed lines "$E$" and "$H$" are defined by 1% of the left peak intensity. Green lines "$F$" and "$G$" are the peak positions of the MP. The horizontal span of each panel is 80°.

**Table 3**
Pulse Width and Polarization Degree of J1518+4904

| | Pulse Width | | $L/I$ | | $V/I$ | |
|---|---|---|---|---|---|---|
| | $W_{10}$(°) | $W_{50}$(°) | ($W_{10}$) | ($W_{50}$) | ($W_{10}$) | ($W_{50}$) |
| OP1 | ⋯ | 14.3 ± 3.7 | ⋯ | 56.8 ± 0.3% | ⋯ | 32.5 ± 0.2% |
| MP | 19.0 | 10.6 | 1.1% | 1.4% | 6.7% | 5.8% |
| PC | >15.1 | >11.3 | 1.6% | 1.6% | 3.1% | 2.6% |
| OP2 | 26.4 ± 3.2 | 11.1 ± 0.5 | 85.5 ± 0.2% | 86.2 ± 0.1% | 8.9 ± 0.1% | 9.4% |

**Note.** We ignore uncertainties below 0.1%.

intensity in the leading component (between $E$ and $F$ as defined in Figure 7) of the MP and the whole MP ($E$ to $H$) and (2) we use the ratio between the integrated intensity in the central component ($F$ to $G$) to that of the whole MP ($E$ to $H$). We also present the statistics of these shape parameters for all the single pulses in the Appendix (Figure 12).

We summed the pulses in different shape groups to form their corresponding profiles and present them in Figure 10. Panel (a) present the pulse profiles of shape groups defined by the leading component ($E$ to $F$) of the MP, while panel (b) presents those of groups defined by the central component ($F$ to $G$) of MP. For panel (a), as $\frac{I_{FG}}{I_{EH}}$ increases: (1) the double peaks of the MP vary from extensive separation to merging. The same shape variation synchronously happens in OP1. (2) The unimodal OP2 shows a consistent expansion of its right side. For panel (b), as $\frac{I_{EF}}{I_{EH}}$ increases (1) the PC shrinks, and its peak transfers from left to right and (2) OP2 shows substructure at its right bottom side when $\frac{I_{EF}}{I_{EH}} < 0.34$. Once the right peak of the MP begins to weaken, the substructure disappears, shown in the red frame with dotted lines.

The analysis shows that there is a correlation between the shapes of the OPs and that of the MP, indicating that they may be coming from the same emission cone but at different heights.

### 3.3. Emission Heights of the Orthogonal Components

Our analysis supports the view that all the OPs we found are from the same pole as their MPs. The emission geometry of PSR J0304+1932 derived from the RVM indicates a high emission height for the OP. The OPs and MP in J1518+4904 display strong strength and shape correlations, suggesting that these components might originate from the same beam as well. In this section we estimate the possible emission heights of the OPs in both pulsars.

The emission altitudes of OPs can be estimated based on pulse widths. In spherical geometry, the angular radius $\rho$ of an emission beam can be derived from its geometric RVM parameters and pulse width $W$, like $\sin^2\left(\frac{W}{4}\right) = \frac{\sin^2(\rho/2) - \sin^2(\beta/2)}{\sin\alpha \sin(\alpha+\beta)}$ (Gil 1981). One can then estimate the emission height by $r_{\rm em} = 2Pc\rho^2/9\pi$ (Kijak & Gil 1997, 1998). For J0304+1932, we take $\alpha = 127°$, $\beta = 3°$, and $W = 130°$. Our estimation indicates that the emission height of J0304+1932 can extend to 0.4 $R_{\rm LC}$, which is much higher than the commonly assumed altitude.

We also estimate the possible altitude by using the phase shift caused by corotated distortion (Blaskiewicz et al. 1991; Dyks & Harding 2004). As part of a corotation frame, the open field lines in the magnetosphere undergo a distortion effect. Such an effect deflects the direction of the radiation beam and hence causes a phase shift of the pulse profile. One can estimate the emission height with the geometrical relationship between the phase shift $\Delta\Phi$ and emission height $r_{\rm em}$ (Shitov 1983; Blaskiewicz et al. 1991; Gangadhara & Gupta 2001; Dyks & Harding 2004; Dyks et al. 2004). This relation has an approximation form like $\Delta\Phi \approx 4r_{\rm em}/R_{\rm LC} - f\sqrt{r_{\rm em}/R_{\rm LC}}$ where parameter $f$ is dependent on the angle between the observation geometry and emission height, within the range $0 < f < 1$ (Dyks & Harding 2004). Assuming $\Delta\Phi$ is a constant parameter, the derivative of the height with respect to $f$ is $\frac{dh}{df} = 1/(2\,h^{-\frac{1}{2}} + \Delta\Phi/2\,h^{-\frac{3}{2}}) > 0$, where $h = r_{em}/R_{LC}$. So we can estimate the lower limit of $r_{em}$ by taking $f = 0$. We use the phase separation between an OP and its MP to refer to the shift of the inflection point of the RVM to the profile peak. For J0304+1932, we measure $\Delta\Phi \sim 100°$ and obtain a height of $r_{em} > 0.44\,R_{LC}$. For J1518+4904, we measure $\Delta\Phi \sim 130°$ for OP1 and 90° for OP2; the estimated heights are $r_{em} > 0.57\,R_{LC}$ for OP1 and $>0.4\,R_{LC}$ for OP2.

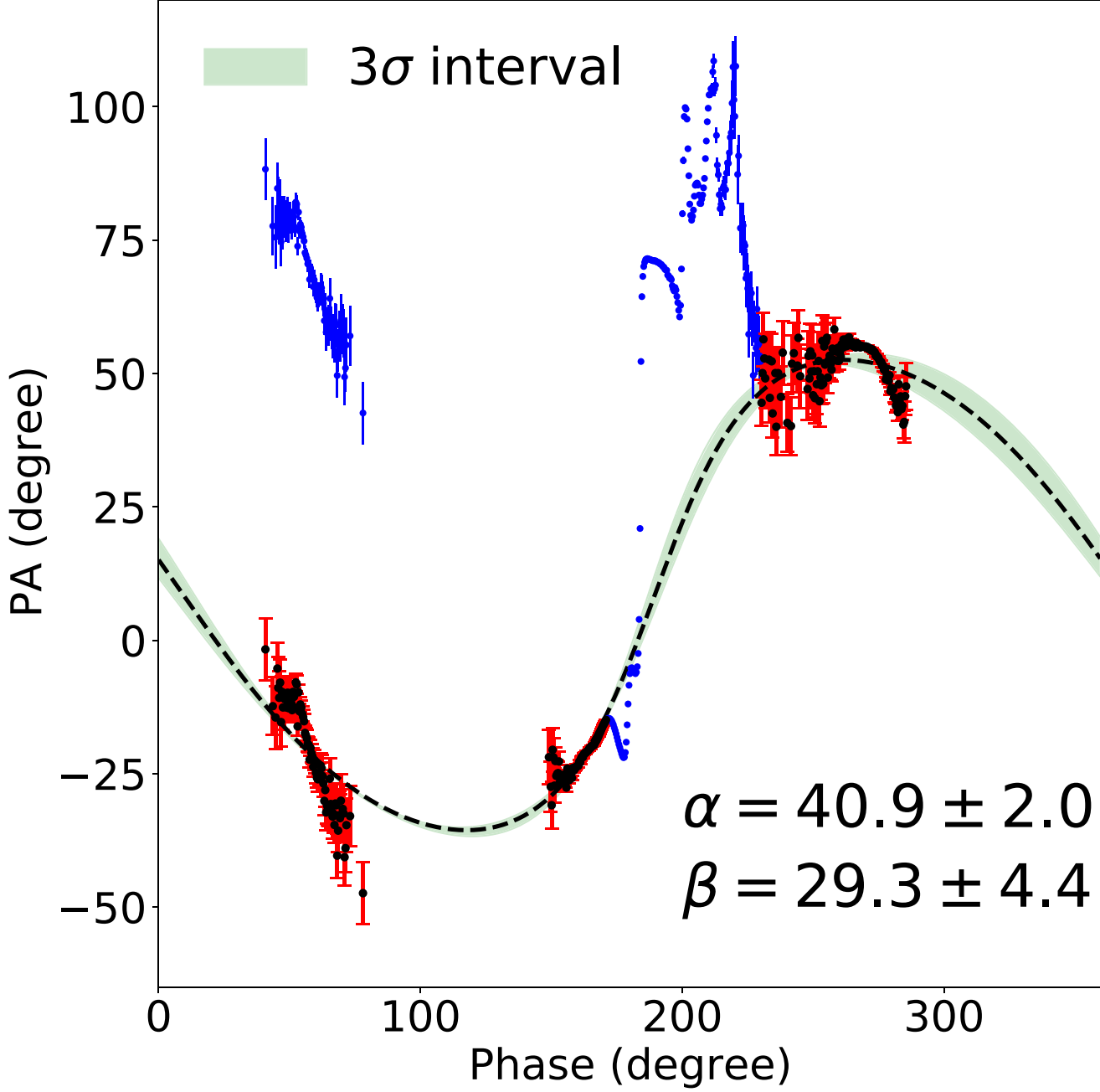

**Figure 8.** An example of an RVM fitting attempt for J1518+4904. We only fit the PA marked with black dots. The PA of the OP (with phase longitude $< 100°$) minus 90°. The dashed black line is the fitting profile. The blue dots (as well as error bars) are PAs we ignored during the fitting. Note that the fit is not a definitive solution because of the complexity of J1518+4904's PA swings. We present this tentative solution here for illustration purposes.

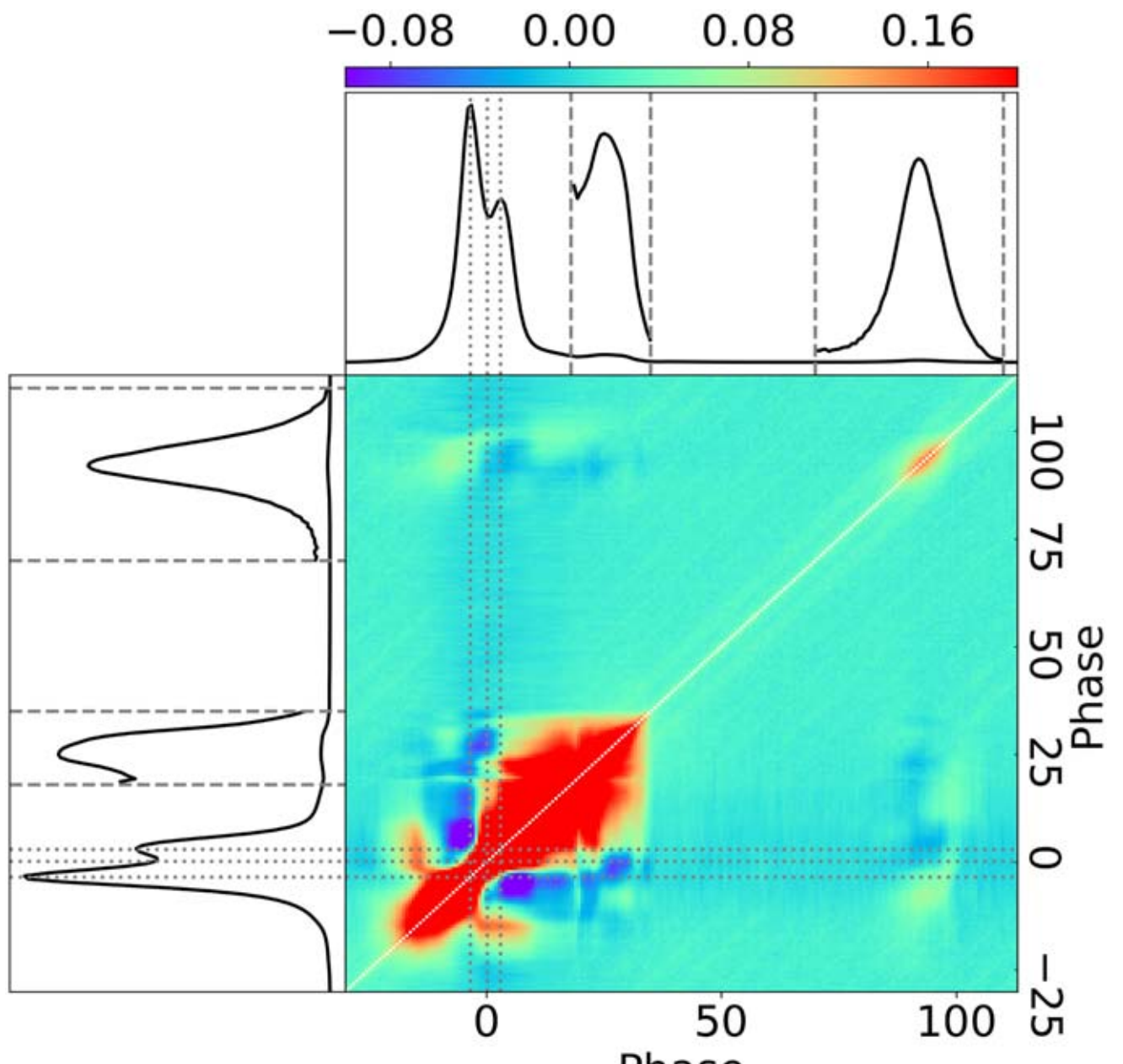

**Figure 9.** The longitude–longitude cross-correlation map of J1518+4904.

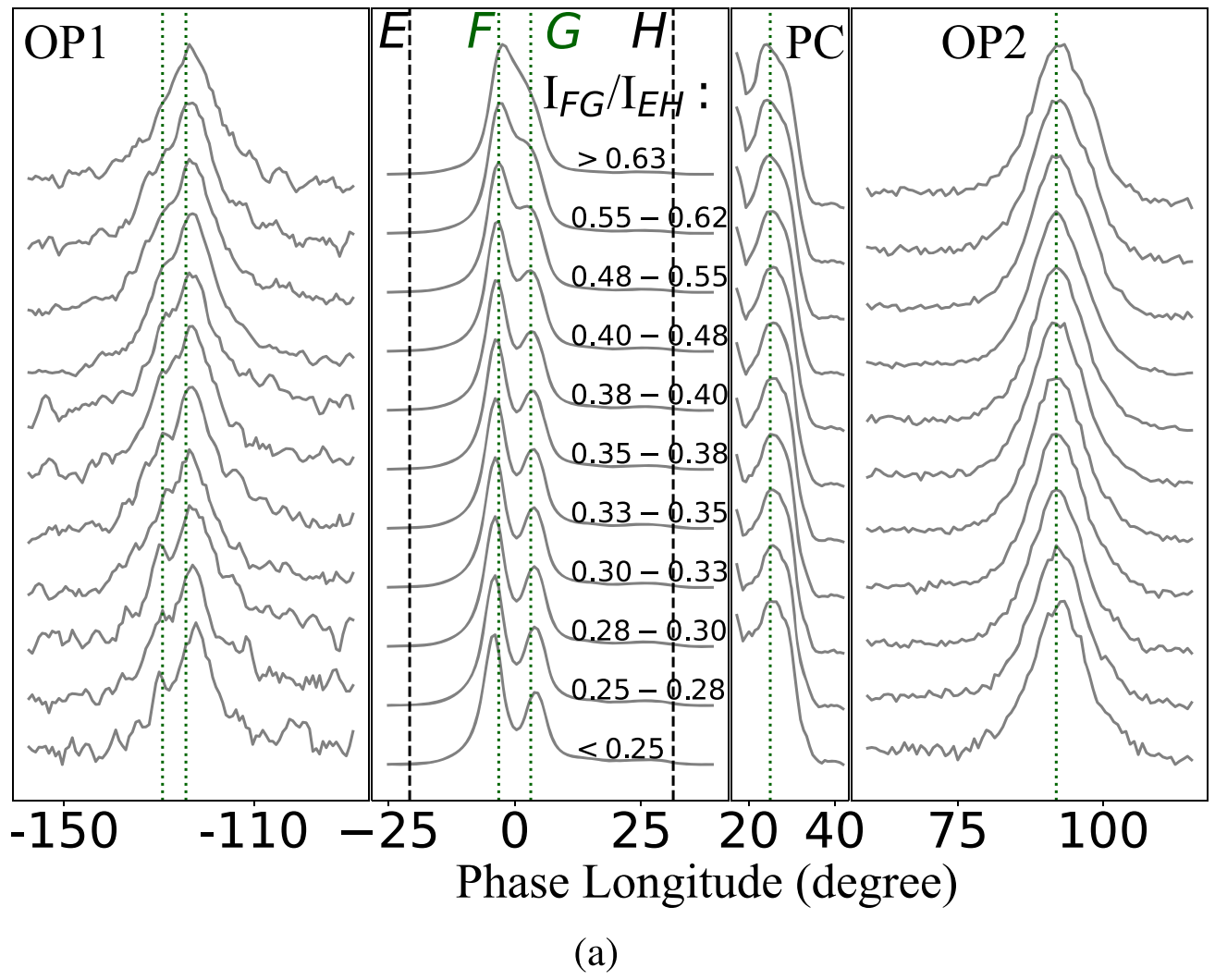

(a)

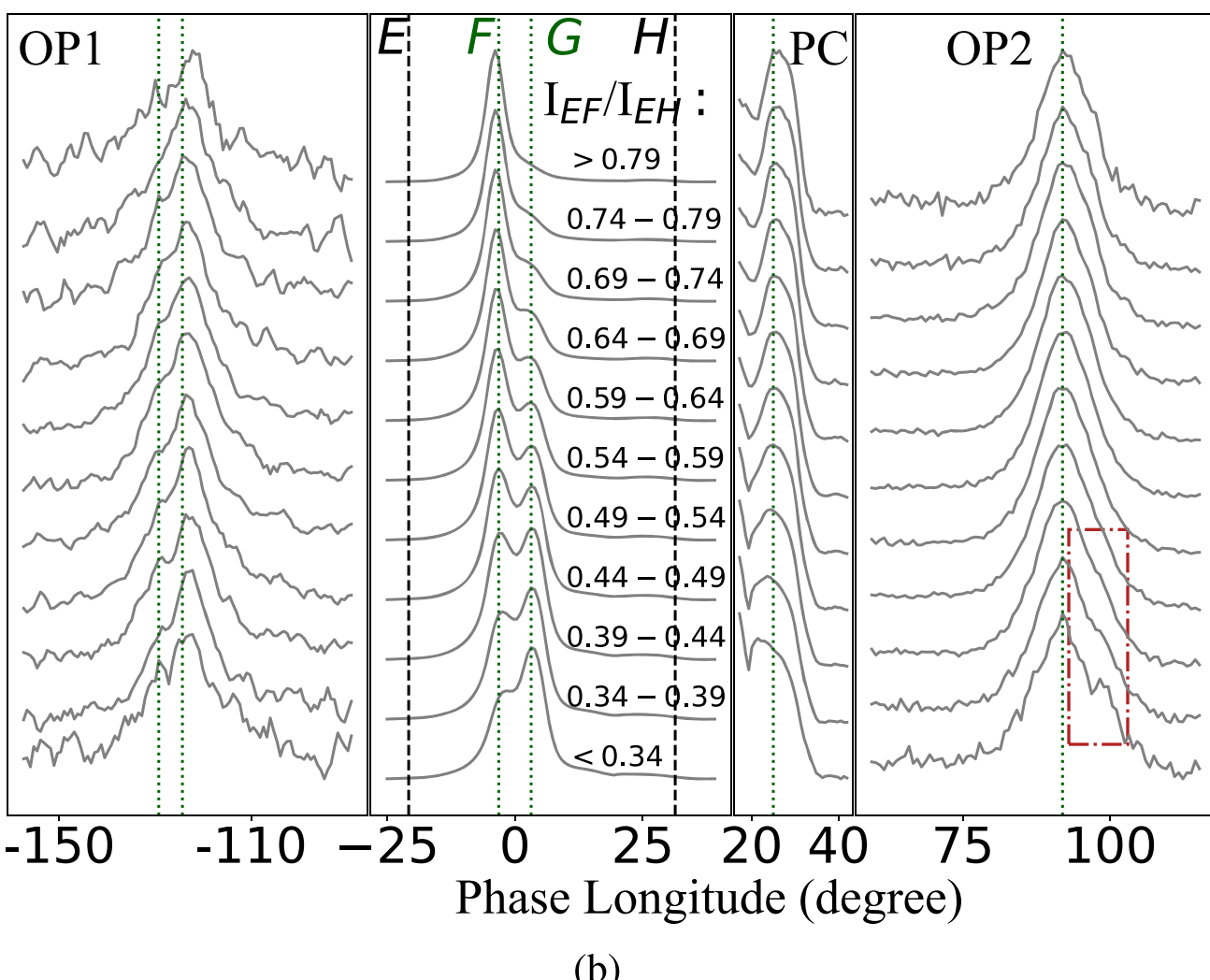

(b)

**Figure 10.** The variation of subintegrated profiles of J1518+4904. Both panels show that as the MP strength varies in the substructures, the OPs behave in correlated variations. Particularly, OP1 totally follows the MP in the profile variation dominated by the central substructure of MP, as panel (a) shows.

A wide open cone may be responsible for such separate high-altitude emissions. It is believed that pulse components from the edge of a beam have higher altitudes than the MP and are usually highly polarized (Lyne & Manchester 1988; Manchester 1995; Karastergiou & Johnston 2007; Johnston et al. 2008; Desvignes et al. 2019). The OPs are consistent with such a scene, especially for J0304+1932, shown in Figures 4 and 5.

We draw a sketch of such a high-altitude emission picture in Figure 11. It naturally explains the OPs distribution of J1518+4904, and the bridge emission between the separated components.

## 4. Discussion

Numerous studies have shown that the radio emission height of pulsars is limited. For nonrecycled pulsars, the altitude is usually within 10% of the light-cylinder radius $R_{LC}$ (Kijak & Gil 1998, 2003; Dyks & Harding 2004; Mitra 2017;

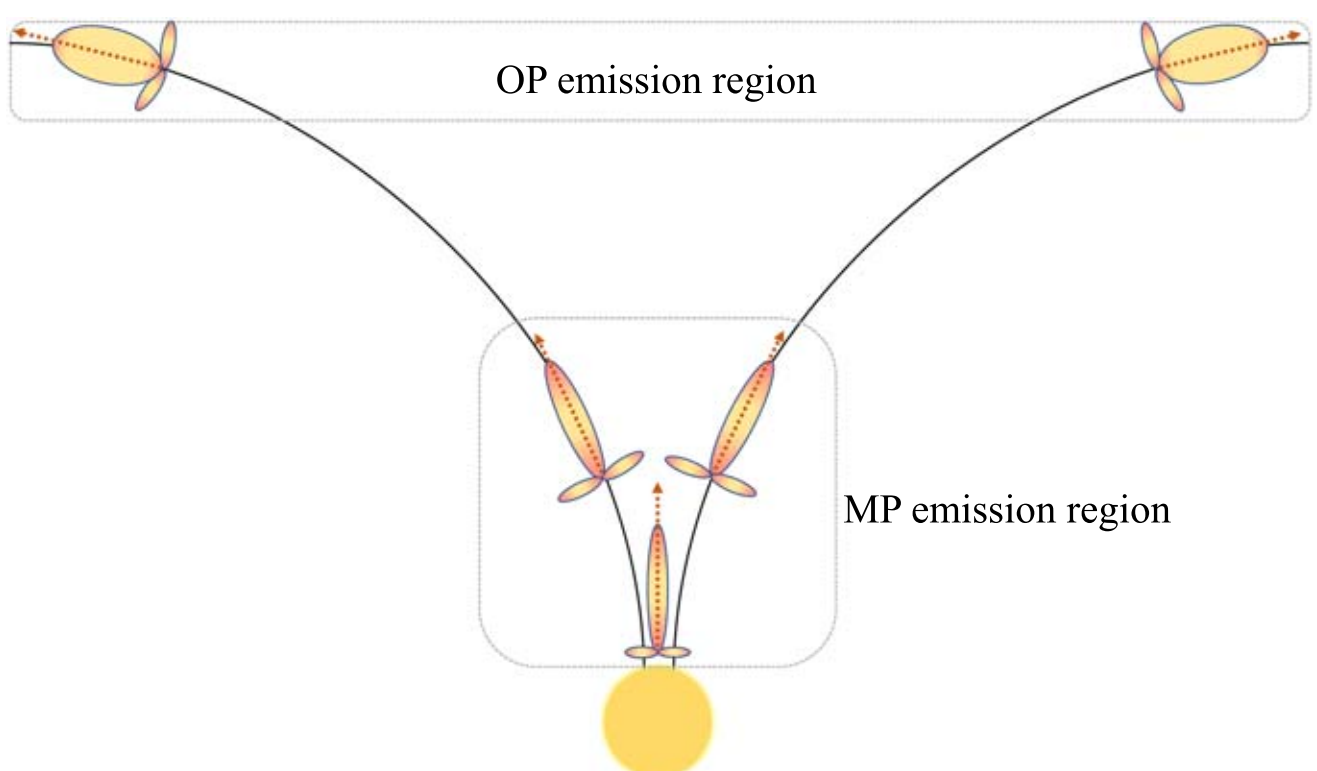

**Figure 11.** Sketch of a pulsar radiation cone where the emissions come from both low and high altitudes in the magnetosphere.

Philippov & Kramer 2022). Some pulsars with emission components more than 10% $R_{\mathrm{LC}}$ may exist (e.g., Basu et al. 2015), but they are still far from the 50% of $R_{\mathrm{LC}}$. For recycled pulsars, they might have higher emission zones with large duty cycles in the phase (Kramer et al. 1998; Gangadhara 2004). The emission from the outer gap zone in a magnetosphere explains the large duty cycle as well as the high-altitude components of recycled pulsars (Xilouris et al. 1998). But these cases are quite different from the OPs in the MP separation longitudes and emission intensity. Our estimations indicate that the separated OPs may originate from a position about half of the light-cylinder radius or even higher. Such emission altitudes lie beyond what is commonly assumed, especially for the nonrecycled pulsar J0304+1932.

Although the high-altitude radiation from the beam edge is accessible to understand the OPs, there are a few caveats to this picture: (1) a complex PA swing of J1518+4904 prevents us from establishing a definitive emission geometry, and the consistent shape change (Figure 10) between the OP1 and MP puzzles us under the beam-edge-emission picture. For J0304+1932, we have analyzed its shape changes in Appendix B. But the OP is not strong enough to show any acceptable morphological correlations to the MP. This result is insufficient to support a same-pole origin of the OP and MP. (2) The correlation between OPs and MP may also be the result of some potential global information communication mechanisms in a pulsar magnetosphere (Biggs 1990; Keith et al. 2010; Latham et al. 2012; Hermsen et al. 2013; Mahajan et al. 2018). For example, the MP of PSR B1055-52 shows a correlation in energy fluctuation with the IP, which is from the opposite pole to the MP (Biggs 1990). (3) Widely separated emission regions lead to an uninterrupted large duty cycle in the phase, which is inconsistent with the separated OPs.

## 5. Conclusion

In this work, we report three separate emission components (OPs) found in two pulsars, J0304+1932 and J1518+4904. Compared to the concept of IPs, all OPs are approximately orthogonal to the MP in phase. Additionally, all of these components are extremely weak and highly linearly polarized. We analyzed the emission geometry of the two pulsars and the correlation between the OPs and their MPs. We also estimated the emission heights of the OPs.

We conclude that the OPs originate from the same magnetic pole as the MPs, but at a higher altitude. The emission heights of the OPs could be half the radius of the light cylinder or even higher.

We would like to thank Michael Kramer for his insightful suggestions and improvements to this paper. This work made use of data from FAST, a Chinese national mega-science facility, built and operated by the National Astronomical Observatories, Chinese Academy of Sciences. This work was funded by National SKA Program of China Nos. 2020SKA0120200, 2020SKA0120100, and 2020SKA0120300 and the National Nature Science Foundation, Nos. 12041303, 11873067, 12041304, and 12003028. This work was also supported by MPG-CAS "LEGAC" collaboration.

## Appendix A
## Statistics of MP Intensity

We present the single-pulse intensity distributions in this section. We take the MP strength, the summed power $I_{AD}$, and $I_{EH}$, as the total intensity of a single pulse for J0304+1932 and J1518+4904, respectively.

Figure 12 shows the intensity statistic of the two pulsars. Here we use the strength ratio of the MP subcomponent to the total intensity (e.g., $I_{AB}/I_{AD}$) to the flux weight of the subcomponent in a single pulse. The power distribution of three subcomponents is presented in gray distribution diagrams. We use the intensity ratio of a single pulse to the mean profile (i.e., $I_{AD}/< I_{AD}$) to measure the strength of a single pulse. Black histogram diagrams show the single-pulse intensity distribution.

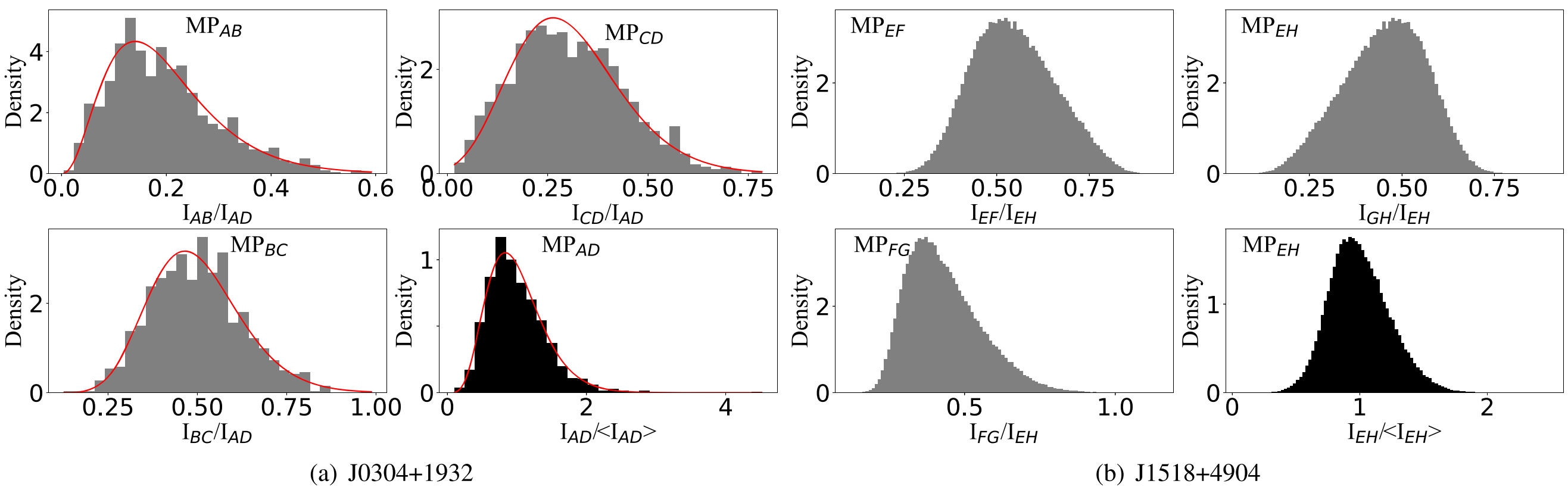

**Figure 12.** The single-pulse intensity statistic. The phase ranges of the different subcomponents are present in Figures 3 and 7. For panel (a), we fit the density with a gamma distribution to get the intensity ratio of the peak density.

## Appendix B
## Shape Changes of PSR J0304+1932

For J0304+1932, we represent the profile variations according to the intensity variation of $MP_{AB}$ and $MP_{BC}$. We divide the single pulses of J0304+1932 into only two subsets due to the relatively low S/N of the OP. Taking $MP_{AB}$ as an example, for a given ratio $i$, we derive two subintegrated profiles. One is integrated by single pulses with $\frac{I_{AB}}{I_{AD}} < i$, and the other profile is the case of $\frac{I_{AB}}{I_{AD}} >= i$. For each substructure, we plot profiles with three given ratios, shown as Figure 13.

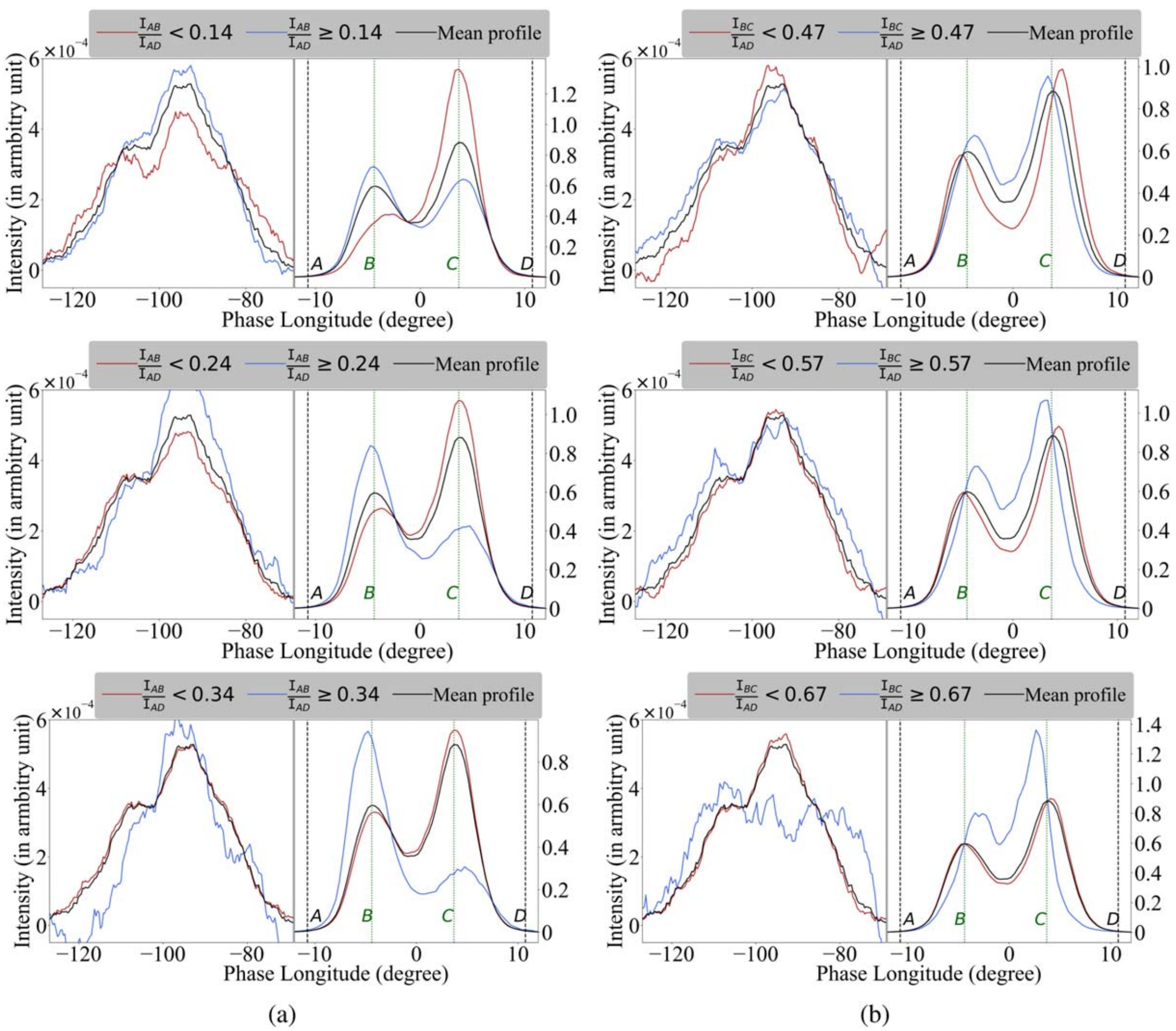

**Figure 13.** The profile variation of J0304+1932. Panel (a) shows the subintegrated profile according to $\frac{I_{AB}}{I_{AD}}$. The red lines from top to bottom are the integrated profiles of single pulses with $\frac{I_{AB}}{I_{AD}} < 0.14$, 0.24, and 0.34. Among them, 0.14 is the intensity value of the peak distribution density, shown in Figure 12. The blue lines are the integrated profile of the remained pulses, respectively. Panel (b) shows the subintegrated pulses according to $\frac{I_{BC}}{I_{AD}}$. The red lines are the integrated profile of single pulses with $\frac{I_{BC}}{I_{AD}} < 0.47$, 0.57, and 0.67 (0.47 is the intensity value of the distribution peaks in Figure 12). The blue lines are the integrated profiles of the remaining pulses, respectively.

## ORCID iDs

Mao Yuan https://orcid.org/0000-0003-1874-0800
Weiwei Zhu https://orcid.org/0000-0001-5105-4058
Michael Kramer https://orcid.org/0000-0002-4175-2271
Bo Peng https://orcid.org/0000-0001-6956-6553
Renxin Xu https://orcid.org/0000-0002-9042-3044
Lijing Shao https://orcid.org/0000-0002-1334-8853
Hong-Guang Wang https://orcid.org/0000-0002-2044-5184
Rushuang Zhao https://orcid.org/0000-0002-1243-0476
Chenchen Miao https://orcid.org/0000-0002-9441-2190
Yi Feng https://orcid.org/0000-0002-0475-7479
Pei Wang https://orcid.org/0000-0002-3386-7159
Di Li https://orcid.org/0000-0003-3010-7661
David J. Champion https://orcid.org/0000-0003-1361-7723
Emmanuel Fonseca https://orcid.org/0000-0001-8384-5049
Jumei Yao https://orcid.org/0000-0002-4997-045X
Paulo C. C. Freire https://orcid.org/0000-0003-1307-9435
Yanjun Guo https://orcid.org/0000-0001-9989-9834